\documentclass[a4paper, 11pt]{article}
\usepackage[pdftex, a4paper]{geometry}
\usepackage{graphicx}				

\usepackage{amsmath}
\usepackage{amsfonts}
\usepackage{dsfont}
\usepackage{mathrsfs}
\usepackage{amssymb}
\usepackage{bbm}
\usepackage{epsfig}
\usepackage[english]{babel}
\usepackage[all]{xy}
\usepackage{color}
\def\T{^{\rm\tiny T}}

\newtheorem{theorem}{Theorem}
\newtheorem{definition}{Definition}

\newtheorem{lemma}{Lemma}
\newtheorem{remark}{Remark}

\begin{document}

\title{\bf Convergence and State Reconstruction of  Time-varying  Multi-agent Systems from Complete Observability Theory}
\author{Brian D. O. Anderson, Guodong Shi,  and Jochen Trumpf\thanks{Authors are  listed in alphabetical order. B. D. O. Anderson is with the National ICT Australia (NICTA),  and the Research School of Engineering, The Australian National University, Canberra, Australia.  G. Shi and J. Trumpf are with the Research School of Engineering, The Australian National University, ACT 0200, Canberra, Australia.    Email:  brian.anderson, guodong.shi, jochen.trumpf@anu.edu.au.}}
\date{}
\maketitle
\begin{abstract}
We study  continuous-time consensus dynamics for multi-agent systems with undirected switching interaction  graphs. We establish a necessary and sufficient condition for exponential asymptotic consensus based on the classical theory of complete observability. The proof is remarkably simple compared to similar results in the literature and the conditions for consensus are mild. This observability-based  method can also be applied to the case where negatively weighted edges are present. Additionally, as a by-product of the observability based arguments, we show that the nodes' initial value can be recovered from the signals on the edges up to a shift of the network average.
\end{abstract}

\section{Introduction}
\subsection{Motivation}
Distributed consensus algorithms   \cite{jad03, xiao04,  saber04,  ren05} of multi-agent systems  have become the foundation of many distributed solutions to in-network  control and estimation \cite{martinez07,kar12},  signal processing \cite{Rabbat2010}, and  optimization  \cite{rabbat2004,nedic09}. Convergence conditions of  consensus dynamics often play a key role in the performance of such distributed designs, which are relatively   well understood under discrete-time network dynamics when the underlying interaction graph is either time-invariant \cite{xiao04,saber04,tsi1} or time-varying \cite{caoming2,tsi2}.

Continuous-time consensus dynamics, however, is difficult to analyze  with switching interaction graphs.  For piece-wise constant graph signals with dwell time, we can treat the continuous-time flow by  investigating the node state evolution along a selected  sequence of time instants  and then work  on the  discrete-time evolution at the sampled times \cite{jad03,ren05}.  For graphs that vary in structure and edge weights in a general manner, the problem becomes much more challenging. Under such circumstances the maximum node state difference  serves as one major tool for convergence analysis \cite{Moreau-2004,lin07,shisiam}.

%

In this paper, we study the convergence of continuous-time consensus dynamics with undirected and switching interaction  graphs from the classical theory on uniformly complete observability of linear time-varying systems \cite{Brian-SIAM-1969,Narendra-SIAM-1977}. It turns out that  the complexity of the convergence analysis can be drastically reduced in view of these classical results even under mild conditions. Interestingly enough, this observability-based  method can also be used to analyze consensus dynamics over networks with both positively and negatively weighted edges \cite{Shi-JSAC-2013,Shi-OR-2016}, where methods based on maximum node state difference function are no longer applicable.
\subsection{The Model}
Consider a multi-agent network with $N$ nodes indexed in the set $\mathbf{V}=\{1,\dots,N\}$. Each node $i$ holds  a state $x_i(t)\in\mathbb{R}$ for $t\in \mathbb{R}^{\geq 0}$. The dynamics of $x_i(t)$ are described by
\begin{align}\label{sys}
\frac{d}{dt}x_i(t)=\sum_{j=1}^N a_{ij}(t) \big( x_j(t)-x_i(t)\big),
\end{align}
where the $a_{ij}(t)=a_{ji}(t)$ are, for the moment, assumed to be  nonnegative numbers representing the weight of the edge between node $i$ and $j$. For the  weight functions $a_{ij}(t)$, we impose the following assumption as a standing assumption throughout our paper.

\medskip

\noindent {\bf Weights Assumption}. (i) Each $a_{ij}(t)$ is regulated in the sense that one-sided limits exist for all $t\geq 0$; (ii) The $a_{ij}(t)$ are almost everywhere bounded, i.e., there is a constant $A^\ast>0$ such that $
|a_{ij}(t)|\leq A^\ast
$
for all $i,j\in\mathbf{V}$ and for almost all $t\geq 0$.
\medskip

  The underlying interaction graph of (\ref{sys}) at time $t$ is defined as an undirected (weighted) graph $\mathbf{G}_t=(\mathbf{V}, \mathbf{E}_t)$, where $\{i,j\}\in\mathbf{E}_t$ if and only if $a_{ij}(t)>0$. We introduce the following definition on the connectivity of the time-varying graph process $\{\mathbf{G}_t\}_{t{\geq 0}}$.

\begin{definition}The time-varying graph $\{\mathbf{G}_t\}_{t{\geq 0}}$ is termed {\it jointly $(\delta,T)$-connected} if there are two real numbers $\delta>0$ and $T> 0$ such that the edges \begin{align}
\{i,j\}:\ \int_{s}^{s+T}a_{ij}(t)dt\geq \delta,\ i,j\in\mathbf{V}
\end{align}
form a connected graph over the node set $\mathbf{V}$ for all $s\geq 0$.
\end{definition}

Along the system (\ref{sys}), it is obvious to see that the average of the node states
$$
x_{\rm ave}(t):=\frac{\sum_{j=1}^N x_j(t)}{N}
$$
is preserved over time. Therefore, if a consensus, i.e., all nodes asymptotically reaching a common state,  is achieved,  then all the node states will converge to $x_{\rm ave}(t_0)$ where $t_0\geq 0$ is the system initial time. Denote by $\mathbf{1}_n$  the $n$-dimensional all one vector and $x(t)=(x_1(t) \cdots x_N(t))\T$. We can therefore conveniently introduce the following definition.

\begin{definition}
Exponential asymptotic  consensus is achieved if there are two constants $\alpha>0$ and $\beta>0$  such that
$$
|x_i(t)-x_{\rm ave}(t_0)|\leq  \beta\|x(t_0)- x_{\rm ave}(t_0)\mathbf{1}_N\| e^{-\alpha (t-t_0)}
$$
for all $i\in\mathbf{V}$, all $x(t_0)$,  and all $t\geq t_0$.
%
\end{definition}

 \subsection{Main Results}

First of all we present a necessary and sufficient condition for exponential asymptotic consensus of the system (\ref{sys}).

 \begin{theorem}\label{theorem1}
Exponential asymptotic consensus is achieved  for the system (\ref{sys}) if and only if there are $\delta>0$ and $T>0$ such that the graph $\{\mathbf{G}_t\}_{t{\geq 0}}$ is  jointly $(\delta,T)$-connected.
\end{theorem}

Theorem \ref{theorem1} is  established in view of the classical  complete observability theory for linear time-varying systems \cite{Brian-TAC-1977,Narendra-SIAM-1977}. A by-product of such observability-based analysis is a state reconstruction theorem, which indicates that node states can be recovered from the signals on the edges up to a shift of the network average. We recall the following  definition.

\begin{definition} (\cite{Brian-SIAM-1969})
A linear time-varying system
\begin{align}
\dot{x}&=F(t)x  \label{1} \\
y&=D\T(t)x \label{2}
\end{align}
is uniformly completely observable if there are some positive $\alpha_1$, $\alpha_2$ and $\delta$ such that
\begin{align}
\alpha_1 I \leq \int_{s}^{s+\delta} \Phi\T(t,s) D(t) D\T(t) \Phi(t,s)dt \leq \alpha_2 I
\end{align}
for all $s\geq 0$. Here $\Phi(\cdot,\cdot)$ is the state transition matrix for  system (\ref{1}).
\end{definition}

If the system (\ref{1}) and (\ref{2}) is uniformly completely observable, then
\begin{align}
&x(s)=\Big( \int_{s}^{s+\delta} \Phi\T(t,s) D(t) D\T(t) \Phi(t,s)dt\Big)^{-1}\nonumber\\
& \int_{s}^{s+\delta} \Phi\T(t,s) D(t) y(t)dt.
\end{align}
In this way,  $x(s)$ is uniformly recovered from the signal $y(t): t\in [s,s+\delta]$.

\begin{theorem}\label{theorem2}Assume  full knowledge of $\{\mathbf{G}_t\}_{t{\geq 0}}$ with the edge weights.
The shifted node state $ x_i(s) -x_{\rm ave}(0)$, $i\in\mathbf{V}
$ can be uniformly recovered from the signals $
 x_j(t) -x_i(t)$, $\{i,j\}\in\mathbf{E}_t$ over the edges
if  there are two constants $\delta>0$ and $T>0$ such that the graph $\{\mathbf{G}_t\}_{t{\geq 0}}$ is jointly $(\delta,T)$-connected.
\end{theorem}

 \subsection{Generalizations}
 We now present a few direct generalizations of the above established convergence results.
 \subsubsection{Robust Consensus}
 Further consider the system (\ref{sys}) subject to noises:
  \begin{align}\label{sys-robust}
\frac{d}{dt}x_i(t)=\sum_{j=1}^N a_{ij}(t) \big( x_j(t)-x_i(t)\big)+w_i(t),\ i\in\mathbf{V},
\end{align}
where $w_i(t)$ is a piecewise continuous function defined over $\mathbb{R}^{\geq 0}$ for each $i\in\mathbf{V}$. We introduce the following notion of robust consensus.

\begin{definition}\label{definition-robust-consensus}
The system (\ref{sys-robust}) achieves  $\ast$-bounded robust consensus if  for all $w(t):=(w_1(t) \dots w_N(t))\T$ satisfying
\begin{align}
\int_{s}^{s+\zeta} w\T(t)w(t) dt\leq B_0
\end{align}
for some $\zeta>0$, some $  B_0 \geq 0$, and all $s\geq 0$, there exists $\mathcal{C}(\zeta,B_0)>0$ such that
\begin{align}
\|x_i(t)-x_{\rm ave}(t)\|\leq \mathcal{C}(\zeta,B_0),\ \forall i\in\mathbf{V}, \forall t\geq t_0
\end{align}
 when $x_1(t_0)=x_2(t_0)=\cdots=x_N(t_0)$.

\end{definition}

The following robust consensus result holds.
 \begin{theorem}\label{theorem3}
The system (\ref{sys-robust}) achieves $\ast$-bounded robust consensus if and only if there are $\delta>0$ and $T>0$ such that the graph $\{\mathbf{G}_t\}_{t{\geq 0}}$ is  jointly $(\delta,T)$-connected.

\end{theorem}
 \subsubsection{Signed Networks}
Up to now we have assumed that all the $a_{ij}(t)$ are non-negative numbers. In fact, the complete observability analysis can be easily extended to the case when some of the $a_{ij}(t)$ are negative. The motivation\footnote{Another definition for a negative link between $i$ and $j$ is to replace $x_j$ by $-x_j$ in the system (\ref{sys}) \cite{Altafini-2013}. The difference of the two definitions of negative links was discussed in \cite{Shi-OR-2016} from a social network point of view.} of studying the effect of possibly  negative $a_{ij}(t)$ comes from the modelling of misbehaved links in engineering network systems, as well as opinion dynamics over  social networks with mistrustful interactions represented by negative edges \cite{Shi-OR-2016}.

Although conventionally, the graph  Laplacian is considered only for  graphs with non-negative edges, for any $t\geq 0$,  we continue to define the (weighted) Laplacian of the graph $\mathbf{G}_t$ as
$$
{L}_{\mathbf{G}_t}= {D}_{\mathbf{G}_t}- {A}_{\mathbf{G}_t}
$$
with $ {A}({\mathbf{G}_t})=[a_{ij}(t)]$ and $ {D}({\mathbf{G}_t})={\rm diag}(\sum_{j=1}^N a_{1j}(t), \dots, \sum_{j=1}^N a_{Nj}(t))$. Note that although the $a_{ij}(t)$ can now be negative,  ${L}_{\mathbf{G}_t}$ remains  symmetric with at least one zero eigenvalue with $\mathbf{1}_N$ being a corresponding  eigenvector. Introduce the following assumption.

\medskip

\noindent {\bf Negative-Link Assumption}. The Laplacian ${L}_{\mathbf{G}_t}$ is positive semi-definite for all $t\geq 0$.

\medskip

This negative-link assumption requires that the influence of the negative links can be reasonably overcome by the positive links for any $t\geq 0$. The following result holds.

\begin{theorem}\label{theorem4}
Consider the system (\ref{sys}) with possibly negative $a_{ij}(t)$.  Assume the Negative-Link Assumption. Then Theorems \ref{theorem1} and \ref{theorem3} continue to hold.
\end{theorem}
  \subsection{Some Remarks}

Theorems \ref{theorem1} and \ref{theorem3} are based  on a direct application of the classical complete observability theory for linear time-varying systems \cite{Brian-TAC-1977,Narendra-SIAM-1977}, after  some useful  properties of the Laplacian of the time-varying network  have been established. The theorems recover the results in  \cite{jad03,Moreau-2004,shisiam} for undirected graphs under  weaker weight and noise assumptions. We would also like to emphasize that the applicability of the complete observability theory relies heavily on undirected node interactions, and  hence we believe that it will be difficult to extend the analysis to the directed case.

Moreover, Theorem \ref{theorem4} illustrates that if the effect of the negative links is somewhat moderated by the positive links, i.e.,  as in the Negative-Link Assumption, then the convergence property is not  affected by the negative links. This point, however, cannot be seen from the max state difference analysis used in prior work  \cite{Moreau-2004,shisiam} because obviously the max state difference function is no longer non-increasing in the presence of negative links, regardless of how small their weights are compared to the positive ones.

Theorem \ref{theorem2} discusses the possibility of reconstructing node initial states from signals over the edges. Naturally one may wonder whether we can recover  the nodes' initial values by observing the state signals at one or more nodes. By duality this is equivalent to the controllability of the network by imposing control inputs over one or more anchor nodes. This problem, as shown in \cite{Rahmani-2009-SIAM,Mesbahi-2013-TAC}, is challenging in general for a full theoretical characterization on general graphs, and in fact,  finding the minimum number of nodes to control to ensure network controllability  is a computationally intensive problem for general network dynamics \cite{Olshevsky-TCNS-2014}.
\section{Proofs of Statements}

\subsection{Proof of Theorem \ref{theorem1}}

Consider the following linear time-varying system
\begin{align}\label{LTV}
\dot{x}=-V(t)V\T(t) x,
\end{align}
where $x\in \mathbb{R}^n$ and $V(\cdot)$ is a piecewise continuous and almost everywhere bounded  matrix function mapping from $\mathbb{R}^{\geq 0}$ to $\mathbb{R}^{n\times r}$. Let $\Phi(\cdot,\cdot)$ be the state transition matrix of the System (\ref{LTV}). Our proof relies on the following result which was proved using complete observability theory in \cite{Brian-TAC-1977}.
\begin{lemma}\label{lem1}
System (\ref{LTV}) is exponentially asymptotically stable in the sense that there are $\gamma_1,\gamma_2>0$ such that $\|\Phi(t,s)\|\leq \gamma_1 e^{-\gamma_2 (t-s)}$ for all $t\geq s\geq 0$,  if and only if for some positive $\alpha_1$, $\alpha_2$ and $\delta$, there holds
\begin{align}\label{100}
\alpha_1 I \leq \int_{s}^{s+\delta} V(t)V\T (t) dt \leq \alpha_2 I
\end{align}
for all $s\in\mathbb{R}^{\geq 0}$.
\end{lemma}

\begin{remark}
 To serve the technical purpose of the proof, we have made a slight variation  in Lemma \ref{lem1} compared to the original  Theorem 1 in \cite{Brian-TAC-1977}, where we have replaced the lower bound on exponentially asymptotic stability required   in \cite{Brian-TAC-1977}  with the almost everywhere boundedness of $V(\cdot)$ adopted in the current paper.
\end{remark}

 The system (\ref{sys}) can be written into the following compact form
\begin{align}\label{laplacian}
\dot{x} =-{L}_{\mathbf{G}_t} x.
\end{align}
Since ${L}_{\mathbf{G}_t}$ is positive semi-definite for all $t\geq 0$ \cite{Godsil-2001}, we can easily construct $V(t)$ so that  the system (\ref{laplacian}) can be re-written into the form of (\ref{LTV}). Although any construction of $V(t)$ would do, we use the incidence matrix $H_{\mathbf{G}_t}$. We make the following definition.
\begin{definition}
The $N\times \frac{N(N-1)}{2}$ weighted incidence matrix of $\mathbf{G}_t$, denoted as  $H_{\mathbf{G}_t}$, with rows indexed by the nodes $v\in \mathbf{V}$ and columns indexed by the edges $\{i,j\}$ (abbreviated as $\{ij\}$), $i<j \in\mathbf{V}$, is defined as $H_{\mathbf{G}_t}=[H_{v-\{ij\}}(t)]$ where\footnote{In this definition we have used an orientation for the graph $\mathbf{G}_t$ in which $i$ is the tail of an edge $\{i,j\}$ if $i<j$. There always holds ${L}_{\mathbf{G}_t}={H}_{\mathbf{G}_t} {H}_{\mathbf{G}_t}\T$ independent of the choice of orientation in the definition of the incidence matrix ${H}_{\mathbf{G}_t}$ \cite{Godsil-2001}.}
\begin{align}
H_{v-\{ij\}}(t)= \left\{ \begin{array}{ll}
         -\sqrt{a_{ij}(t)}, & \mbox{if $v=i$};\\
          \sqrt{a_{ij}(t)}, & \mbox{if $v=j$};\\
        0, & \mbox{otherwise}.\\\end{array} \right.
\end{align}
\end{definition}

We can now write (\ref{laplacian}) as
\begin{align}\label{11}
\dot{x} =-{L}_{\mathbf{G}_t} x= -H_{\mathbf{G}_t} H_{\mathbf{G}_t}\T x.
\end{align}

 Define $y(t)=(I-\mathbf{1}_N\mathbf{1}_N\T/N)x(t)$. Then we have
\begin{align}\label{3}
\dot{y}&=(I-\mathbf{1}_N\mathbf{1}_N\T/N)\dot{x}(t)\nonumber\\
&=-(I-\mathbf{1}_N\mathbf{1}_N\T/N)L_{\mathbf{G}_t}  x(t)\nonumber\\
&\stackrel{a)}{=}-(L_{\mathbf{G}_t}+\mathbf{1}_N\mathbf{1}_N\T/N)(I-\mathbf{1}_N\mathbf{1}_N\T/N)  x(t)\nonumber\\
&= -(L_{\mathbf{G}_t}+\mathbf{1}_N\mathbf{1}_N\T/N) y(t)\nonumber\\
&\stackrel{b)}{=} -\Big(H_{\mathbf{G}_t}+ \frac{  \mathbf{1}_N \mathbf{1}_r\T}{\sqrt{rN}} \Big) \Big(H_{\mathbf{G}_t}+\frac{  \mathbf{1}_N \mathbf{1}_r\T}{\sqrt{rN}} \Big)\T y(t),
\end{align}
where $r={N(N-1)}/{2}$, $a)$  follows from the equations  that $(I-\mathbf{1}_N\mathbf{1}_N\T/N)L_{\mathbf{G}_t}=L_{\mathbf{G}_t}(I-\mathbf{1}_N\mathbf{1}_N\T/N)$ and $(\mathbf{1}_N\mathbf{1}_N\T/N)^2=\mathbf{1}_N\mathbf{1}_N\T/N$, and $b)$ follows from direct computation in view of ${L}_{\mathbf{G}_t}={H}_{\mathbf{G}_t} {H}_{\mathbf{G}_t}\T$.

We can now conclude that the following statements are equivalent.

\begin{itemize}
\item[(i)] System (\ref{laplacian}) achieves exponential consensus;

\item[(ii)] System (\ref{3}) is exponentially asymptotically stable;

\item[(iii)] There are some positive $\alpha_1$, $\alpha_2$ and $T$, such that
\begin{align}\label{10}
\alpha_1 I &\leq \int_{s}^{s+T}\Big(H_{\mathbf{G}_t}+ \frac{  \mathbf{1}_N \mathbf{1}_r\T}{\sqrt{rN}} \Big) \Big(H_{\mathbf{G}_t}+\frac{  \mathbf{1}_N \mathbf{1}_r\T}{\sqrt{rN}} \Big)\T dt\nonumber\\
&=\int_{s}^{s+T}(L_{\mathbf{G}_t}+\mathbf{1}_N\mathbf{1}_N\T/N) dt \leq \alpha_2 I;
\end{align}
for all $s\in\mathbb{R}^{\geq 0}$.

\item[(iv)]  there are $\delta, T>0$ such that the graph $\{\mathbf{G}_t\}_{t\geq 0}$ is  $(\delta,T)$-connected.
\end{itemize}

The equivalence of (i) and (ii) is straightforward. Lemma \ref{lem1} gives the equivalence between (ii) and (iii). We only need to show the equivalence between (iii) and (iv).

\noindent (iii) $\Rightarrow$ (iv): If (\ref{10}) holds, then $\int_{s}^{s+T}(L_{\mathbf{G}_t}+\mathbf{1}_N\mathbf{1}_N\T/N) dt $ is positive definite. Note that $\mathbf{1}_N\mathbf{1}_N\T/N$ has a unique positive eigenvalue equal to $1$ and another eigenvalue equal to zero with multiplicity  $N-1$. Moreover, for any fixed $t\geq0$, $L_{\mathbf{G}_t} \mathbf{1}_N\mathbf{1}_N\T/N =\big(\mathbf{1}_N\mathbf{1}_N\T/N\big)L_{\mathbf{G}_t}=0$, and the zero eigenvalue of  $L_{\mathbf{G}_t}$ and the eigenvalue one of  $\mathbf{1}_N\mathbf{1}_N\T/N$ share a common eigenvector $\mathbf{1}_N$. This implies that the second smallest eigenvalue of $\int_{s}^{s+T}L_{\mathbf{G}_t}dt$, $\lambda_2(\int_{s}^{s+T}L_{\mathbf{G}_t}dt)$, must satisfy
$\lambda_2(\int_{s}^{s+T}L_{\mathbf{G}_t}dt)\geq \alpha_1$.

Consider $\lambda_2(L)$, the second smallest eigenvalue of $L$, where $L$ takes value from
\begin{align*}
&\mathscr{L}:=\big\{L\in \mathbb{R}^{N\times N}: L\ \mbox{is symmetric}; [L]_{ij}\leq 0, i\neq j; \\
&\ \ \ L\mathbf{1}_n=0 \big\}.
\end{align*}
Certainly $\lambda_2(\cdot)$ is a continuous function over the set $\mathscr{L}$.  In turn, there must exist $\delta>0$ (which depends on $\alpha_1$)  such that the edges \begin{align}
\{i,j\}:\ \int_{s}^{s+T}a_{ij}(t)dt\geq \delta,\ i,j\in\mathbf{V}
\end{align}
form a connected graph over the node set $\mathbf{V}$. In other words, the graph $\{\mathbf{G}_t\}_{t\geq 0}$ is  $(\delta,T)$-connected.

\noindent (iv) $\Rightarrow$ (iii):   The matrices $\int_{s}^{s+T}L_{\mathbf{G}_t}dt$, for all $s\geq0$ and all $(\delta,T)$-connected graphs $L_{\mathbf{G}_t}$, form a compact set in $\mathscr{L}$. Therefore, again noticing that $\lambda_2(\cdot)$ is a continuous function over the set $\mathscr{L}$,  we can find a constant $\alpha_1>0$ (which depends on $\delta$) such that $\lambda_2(\int_{s}^{s+T}L_{\mathbf{G}_t}dt)\geq \alpha_1$ for all $s\geq 0$ and for all $(\delta,T)$-connected graph $\{\mathbf{G}_t\}_{t\geq 0}$. Re-using the above properties of the two matrices $L_{\mathbf{G}_t}$ and $\mathbf{1}_N\mathbf{1}_N\T/N$ we immediately get the lower bound part of (\ref{10}). The existence of $\alpha_2$ in (\ref{10}) is always guaranteed by the Weights Assumption on the almost everywhere boundedness of the $a_{ij}(t)$.

We have now proved Theorem \ref{theorem1} by the equivalence of (i) and (iv).

\subsection{Proof of Theorem \ref{theorem2}}
From the proof  of Theorem \ref{theorem1}, if there are $\delta, T>0$ such that the graph $\mathbf{G}(t)$ is  $(\delta,T)$-connected, then
there are some positive $\alpha_1$, $\alpha_2$ and $T$, there holds
\begin{align}
\alpha_1 I \leq\int_{s}^{s+T}\Big(H_{\mathbf{G}_t}+ \frac{  \mathbf{1}_N \mathbf{1}_r\T}{\sqrt{rN}} \Big) \Big(H_{\mathbf{G}_t}+\frac{  \mathbf{1}_N \mathbf{1}_r\T}{\sqrt{rN}} \Big)\T dt\leq \alpha_2 I;
\end{align}
for all $s\geq 0$. This in turn implies that the following system (see \cite{Brian-TAC-1977})
\begin{align}
\left\{ \begin{array}{ll}
        \dot{y}&= -(L_{\mathbf{G}_t}+\mathbf{1}_N\mathbf{1}_N\T/N) y;\\
   z&=\Big(H_{\mathbf{G}_t}+\frac{ \mathbf{1}_N \mathbf{1}_r\T}{\sqrt{rN}} \Big)\T y. \end{array} \right.
\end{align}
is uniformly completely observable.

Note that
\begin{align}
z(t)&=\Big(H_{\mathbf{G}_t}+\frac{ \mathbf{1}_N \mathbf{1}_r\T}{\sqrt{rN}} \Big)\T y(t)\nonumber\\
&=\Big(H_{\mathbf{G}_t}\T+\frac{ \mathbf{1}_r \mathbf{1}_N\T}{\sqrt{rN}} \Big)(I-\mathbf{1}_N\mathbf{1}_N\T/N){x}(t)\nonumber\\
&=H_{\mathbf{G}_t}\T{x}(t).
\end{align}
This immediately translates to the desired theorem.
\subsection{Proof of Theorem \ref{theorem3}}
Again we consider $y(t)=(I-\mathbf{1}_N\mathbf{1}_N\T/N)x(t)$. Along the system (\ref{sys-robust}) the evolution of $y(t)$ obeys
\begin{align}\label{4}
\frac{d}{dt}y(t) = -(L_{\mathbf{G}_t}+\mathbf{1}_N\mathbf{1}_N\T/N) y(t)+(I-\mathbf{1}_N\mathbf{1}_N\T/N) w(t).
\end{align}
It is then obvious that  system (\ref{sys-robust}) achieves  $\ast$-bounded robust consensus under Definition \ref{definition-robust-consensus} if system (\ref{4}) is bounded*-input, bounded state stable as defined in \cite{Brian-SIAM-1969}.

If  the graph $\{\mathbf{G}_t\}_{t\geq 0}$ is  $(\delta,T)$-connected, then (\ref{3}) is exponentially asymptotically stable. This in turn leads to the conclusion that the system (\ref{4}) is bounded*-input, bounded state stable based on the sufficiency part of Theorem 1 (p. 404) in \cite{Brian-SIAM-1969}. Therefore,  system (\ref{sys-robust}) achieves  $\ast$-bounded robust consensus.

On the other hand, note that there holds
\begin{align}
(L_{\mathbf{G}_t}+\mathbf{1}_N\mathbf{1}_N\T/N)(I-\mathbf{1}_N\mathbf{1}_N\T/N)=(L_{\mathbf{G}_t}+\mathbf{1}_N\mathbf{1}_N\T/N)
\end{align}
for all $t$. Consequently, we have
\begin{align}
\Phi(t,s)(I-\mathbf{1}_N\mathbf{1}_N\T/N)=\Phi(t,s)
\end{align}
for all $t$ and $s$, where $\Phi(t,s)$ is the state-transition matrix of the system (\ref{4}) with $w(t)\equiv 0$. Therefore, applying  the necessity proof of Theorem 1 in \cite{Brian-SIAM-1969} we can directly conclude that if system (\ref{4}) is bounded*-input, bounded state stable, then (\ref{3}) is exponentially asymptotically stable. This in turn leads to  the graph $\{\mathbf{G}_t\}_{t\geq 0}$ being  $(\delta,T)$-connected.

We have now completed the proof.
\subsection{Proof of Theorem \ref{theorem4}}
When the Negative-Link Assumption holds, the construction of (\ref{11}) can be replaced by
\begin{align}
\dot{x} =-{L}_{\mathbf{G}_t} x= -\sqrt{{L}_{\mathbf{G}_t}} \sqrt{{L}_{\mathbf{G}_t}} x.
\end{align}
It is then straightforward to see that the remaining  arguments in the proofs of Theorems \ref{theorem1} and \ref{theorem3} will not be affected, leading to the desired result.

\section{Conclusions}
We have studied continuous-time consensus dynamics  with undirected switching interaction graphs in view of the classical theory of uniform complete observability  for linear time-varying systems. A necessary and sufficient condition for exponential asymptotic consensus was established by much simplified analysis   compared to related results in the literature. Interestingly and importantly, this observability-based  method can also be applied to signed networks with  both positive and negative edges. We also established a robust consensus result as well as a state reconstruction result indicating  that the nodes' initial values can be recovered from signals on the edges up to  a shift of the network average.

\end{document}